\documentclass{article}

\usepackage{arxiv}

\usepackage[utf8]{inputenc} 
\usepackage[T1]{fontenc}    
\usepackage{hyperref}       
\usepackage{cleveref}
\usepackage{multirow}
\usepackage{url}            
\usepackage{booktabs}       
\usepackage{amsfonts}       
\usepackage{nicefrac}       
\usepackage{microtype}      
\usepackage{lipsum}
\usepackage{fancyhdr}       
\usepackage{graphicx}       
\graphicspath{{media/}}     

\title{Interactive Generation of Laparoscopic Videos with Diffusion Models
}

\author{
Ivan Iliash\\
Technical University of Munich \\
\texttt{ivan.iliash@gmail.com} \\
\AND 
\hspace{-15pt}Simeon Allmendinger \\
\hspace{-15pt}University of Bayreuth\\
\hspace{-15pt}\texttt{simeon.allmendinger@uni-bayreuth.de}
\And
\hspace{-15pt} Felix Meissen \\
\hspace{-15pt}Technical University of Munich\\
\hspace{-15pt}\texttt{felix.meissen@tum.de} \\
\AND
\hspace{20pt} Niklas Kühl \\
\hspace{20pt} University of Bayreuth\\
\hspace{30pt} \texttt{kuehl@uni-bayreuth.de} \\
\And
\hspace{16pt} Daniel Rückert\\
\hspace{16pt} Technical University of Munich\\
\hspace{16pt} \texttt{daniel.rueckert@tum.de} \\
}

\begin{document}
\maketitle

\begin{abstract}
Generative AI, in general, and synthetic visual data generation, in specific, hold much promise for benefiting surgical training by providing photorealism to simulation environments. Current training methods primarily rely on reading materials and observing live surgeries, which can be time-consuming and impractical. In this work, we take a significant step towards improving the training process. Specifically, we use diffusion models in combination with a zero-shot video diffusion method to interactively generate realistic laparoscopic images and videos by specifying a surgical action through text and guiding the generation with tool positions through segmentation masks.
We demonstrate the performance of our approach using the publicly available Cholec dataset family and evaluate the fidelity and factual correctness of our generated images using a surgical action recognition model as well as the pixel-wise F1-score for the spatial control of tool generation. We achieve an FID of $38.097$ and an F1-score of $0.71$.


\keywords{Diffusion Models \and Video Generation \and Laparoscopic Surgery}
\end{abstract}
\section{Introduction}
Surgical simulations offer a significant advantage by eliminating the need for patient involvement in skills practice, providing trainees with essential technical lessons before performing procedures on humans \cite{sutherland2006_surgicalsimulation}.
However, current computer-based simulations have lots of drawbacks, such as unrealistic visual appearance, lacking variability, and complex creation procedures taking into account the varying anatomical properties, all of which lead to diminishing the quality of surgical training.
Therefore, AI-generated surgical simulations promise significant advancements in medical education since the underlying machine-learning models can learn the anatomical and visual characteristics of surgeries as well as their interactions with surgical tools from real-world data. Similar to recent works on image-guided surgery by Ramalhinho \textit{et al.} \cite{imageguided2} and Schneider \textit{et al.} \cite{imageguided3}, our work focuses on laparoscopic surgery.

We propose an approach for generating realistic laparoscopic videos conditioned on both text prompts and surgical tool positions. This lays the groundwork for a dynamic and interactive surgical training platform that mimics real-world scenarios. With this approach, we achieve state-of-the-art realism with an FID score of $33.43$ and a pixel-wise F1 score of $0.72$ for the control of tool positions. Moreover, we successfully generate coherent videos of single surgical actions. Our contribution represents, to our knowledge, the first entry in diffusion-based controllable surgical video generation. Exemplary results and code are made accessible on our project page~\footnote{\href{https://anonymous.4open.science/w/laparoscopic-video-generation-D1C3/}{https://anonymous.4open.science/w/laparoscopic-video-generation-D1C3/}}.

\section{Related Work}
\subsubsection{Medical Image Generation} has been an area of interest for many years now. However, well-known diffusion models, such as OpenAI's DALL-E 3~\footnote{\href{https://cdn.openai.com/papers/dall-e-3.pdf}{https://cdn.openai.com/papers/dall-e-3.pdf}} model struggle to create meaningful laparoscopic images. Thus, research focuses on generative models specified to this particular medical domain. Often, image-to-image methods are used in which the conditioning image contains the structural information for guidance. Pfeiffer \textit{et al.}~\cite{laparoscopy_img2img} use Generative Adversarial Networks (GANs) for laparoscopic image-to-image translation, going from simple 3D renders to photorealistic scenes, which enables more easily obtainable 3D data. Kaleta \textit{et al.} \cite{sd_cn_endoscopic} pursued a similar goal by combining ControlNet \cite{controlnet} with a StableDiffusion model finetuned with Dreambooth \cite{ruiz2023dreambooth}.
Further research makes use of text-to-image models to generate chest-Xrays~\cite{chambon2022roentgen} and laparoscopic images~\cite{allmendinger2023navigating}, while others use alternative conditioning based on class labels, e.g., surgical phase and tool presence~\cite{cataract_diffusion}.

\subsubsection{Medical Video Generation} is a relatively unexplored topic due to its novelty.
Recent work builds upon a video diffusion model that generates echocardiogram sequences starting from a real image as the first frame and left ventricular ejection fraction conditioning~\cite{reynaud2023feature}. Another method uses diffusion models to generate cardiac magnetic resonance imaging videos by combining a deformation module with a 3D-UNet architecture \cite{kim2022diffusion}.

\section{Video Generation Pipeline}

\begin{figure}[t]
    \centering 
    \includegraphics[scale=.43]{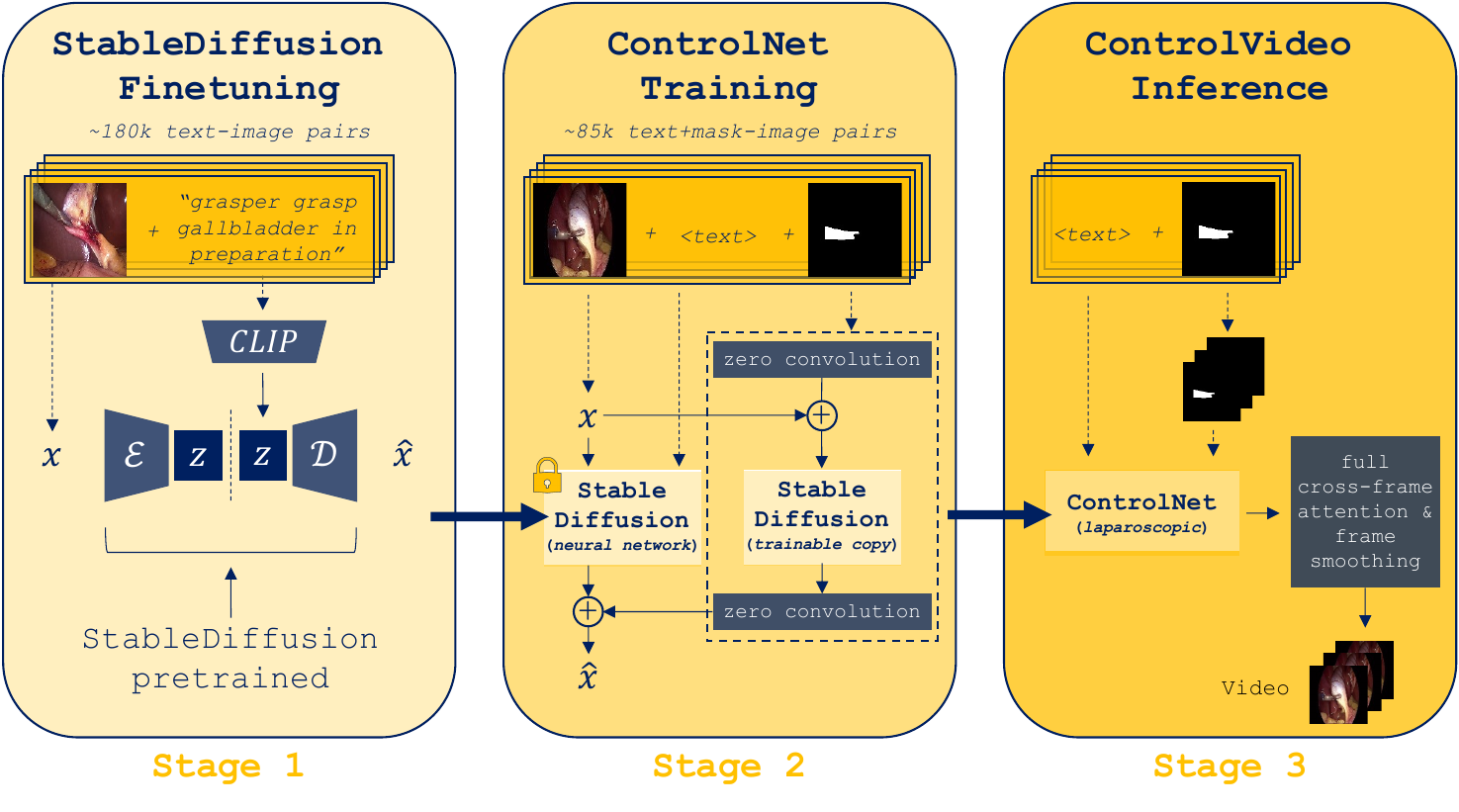}
    \caption{Overview of our approach.
    Consisting of three stages: \texttt{StableDiffusion} \textbf{finetuning} to adapt the text-to-image model to the laparoscopic domain. \texttt{ControlNet} \textbf{training} for adding conditioning with tool segmentation masks. \texttt{ControlVideo} \textbf{inference} for controlled video generation.}
    \label{fig:overview}
\end{figure}

Our approach consists of three stages, which build a training and inference pipeline to generate laparoscopic videos. \Cref{fig:overview} displays the individual stages and their relation.\\\\
\textbf{Stage 1: \texttt{StableDiffusion} \cite{stablediffusion} Finetuning}. The basis of our approach is created by adapting a popular text-to-image model to generate laparoscopic images using surgical action descriptions as prompts. \\
\textbf{Stage 2: \texttt{ControlNet} \cite{controlnet} Training}. Building upon the finetuned StableDiffusion model, we train an extension for it to generate images conditioned on surgical tool segmentation masks in addition to text prompts. \\
\textbf{Stage 3: \texttt{ControlVideo} \cite{controlvideo} Inference}. We utilize a zero-shot extension to ControlNet that ensures consistency between generated laparoscopic images in subsequent video frames.

\subsection{StableDiffusion}
Denoising Diffusion Probabilistic Models (DDPMs)\cite{ho_ddpm} are a class of generative models that learn to generate data by reversing a diffusion process. The diffusion process gradually adds noise to the data over a series of steps, transforming the original data distribution into a known noise distribution (usually Gaussian). A UNet~\cite{ronneberger_unet} model is then trained to learn the reverse process, denoising the data step by step to recover the original data distribution from the noise.
StableDiffusion~\cite{stablediffusion} is a latent diffusion model architecture built upon DDPMs, which integrates a variational autoencoder (VAE) to compress high-dimensional inputs into a more manageable lower-dimensional latent space, facilitating efficient training and processing. The incorporation of a CLIP~\cite{clip} text encoder for conditioning transforms text prompts into semantic embeddings, guiding the image generation to closely reflect the content described in the text.
In Stage 1, we finetune the weights of a pre-trained model\footnote{\href{https://huggingface.co/runwayml/stable-diffusion-v1-5}{https://huggingface.co/runwayml/stable-diffusion-v1-5}}{} using a dataset of laparoscopic text-image pairs. 

\subsection{ControlNet}
For the purpose of enabling spatial conditioning, ControlNet~\cite{controlnet} augments the StableDiffusion framework beyond the conventional text-to-image architecture. This augmentation involves freezing the parameters of StableDiffusion's UNet and adding trainable copies of its encoder blocks. The trainable copies are connected to the original UNet via \textit{zero convolutions}, which are $1\times1$ convolutional layers with both their weights and biases initialized to zero. This approach safeguards the pre-trained model against adverse effects of noise during the initial stages of training, enabling a progressive accumulation of desired characteristics while preserving previously acquired capabilities. 

In Stage 2, ControlNet is utilized to precisely manipulate the positioning of surgical tools in synthetic images by employing segmentation masks. We limit ourselves to using only tool masks as conditioning since this information can be easily obtained or simulated. We rely on the diffusion model to generate the rest of the environment to enable a bigger variability in the synthetic data, which also reflects the differences between each patient.

\subsection{ControlVideo}
ControlVideo \cite{controlvideo} is a zero-shot framework that extends ControlNet by full cross-frame attention to generate temporally coherent video sequences and an interleaved frame smoother to reduce flickering. This enhancement involves transforming the StableDiffusion UNet's 2D convolution layers ($k$$\times$$k$) into 3D convolutions ($1$$\times$$k$$\times$$k$), with the new temporal dimension indexing video frames. The self-attention blocks are then replaced with cross-attention over all frames. Consequently, in Stage 3, we provide a text prompt along with subsequent conditioning frames of tool segmentation masks.

\section{Experiments}
To validate our approach, we conduct technical experiments using publicly available datasets. 

\begin{figure}[h]\
    \centering
    \makebox{\includegraphics[width=1\textwidth]{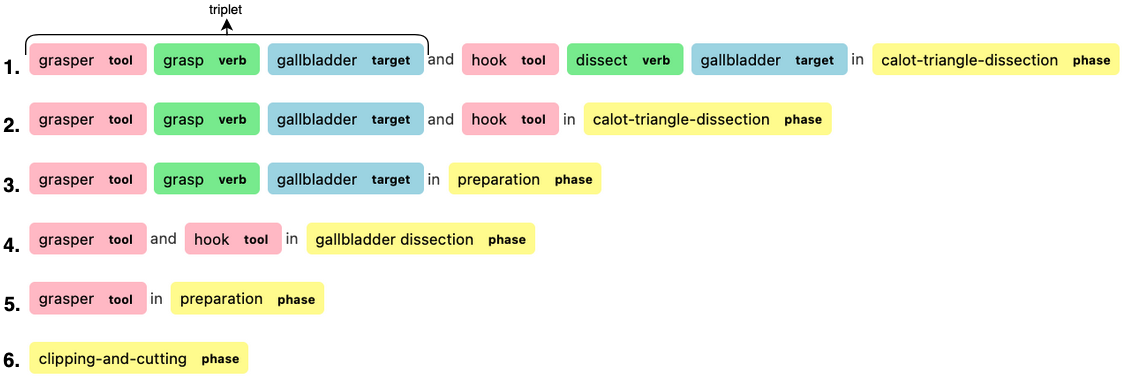}}
    \caption{Exemplary prompts used to condition the text-to-image model. The formats from the first three examples are only available in the CholecT50 dataset, as verb and target labels are missing in the Cholec80 dataset.}
    \label{fig:prompts}
\end{figure}

\subsection{Data}
\textbf{CholecT50} \cite{cholect50} is a collection of 50 laparoscopic cholecystectomy surgery videos with a total of 100,863 frames. Each frame is labeled with surgical action text triplets in the format $<$\textit{tool, verb, target}$>$. In total, there are 6 tools, 8 verbs, 14 target classes, and 101 realistically possible triplets. Additionally, labels for the surgical tool's presence and the current surgical phase are included. \textbf{Cholec80} \cite{cholec80} is a superset of CholecT50 comprised of 80 videos and solely annotated with labels of surgical phases and tool presence. Thus, the dataset is less informational but increases the number of frames by 94,056 additional ones.
We resize all images to 128$\times$128 and extract text prompts from both datasets with changing formats depending on the available labels exemplary shown in \Cref{fig:prompts}.

For StableDiffusion finetuning, we employ two different dataset variants: The first variant solely consists of accurate triplet labels from CholecT50. The second variant also contains the remaining Cholec80 frames. We differentiate to evaluate the effect of the less accurate labels, especially on the factual correctness of the resulting generated images.

\begin{figure}[t]
    a) \textbf{Actual images drawn from the CholecT50 dataset.}\\
    \includegraphics[width=1\textwidth]{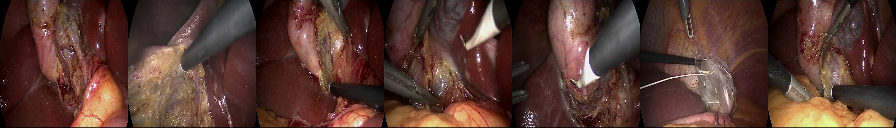}\\
    b) \textbf{Generated images applying the finetuned StableDiffusion model.}\\
    \includegraphics[width=1\textwidth]{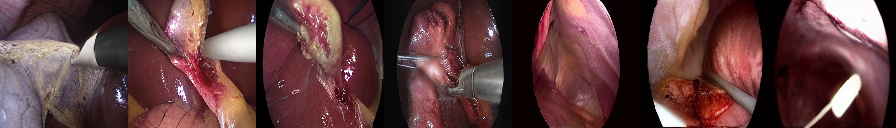}
    \caption{Actual images from the CholecT50 dataset (a) vs. images generated by our finetuned StableDiffusion model (b).}
    \label{fig:examples}
    \vspace{-15px}
\end{figure}

As there is no tool segmentation mask for the majority of the frames over all \textit{Cholec} datasets, we use two publicly available laparoscopic tool segmentation datasets to fine-tune a YOLOv8 \cite{yolov8} model:
The \textbf{CholecSeg8k} \cite{cholecseg8k} dataset, which contains 8000 semantically segmented images from the Cholec80 and the \textbf{ART-Net} \cite{artnetseg} dataset, which consists of 1324 laparoscopic images with tool segmentation masks.
We use our finetuned segmentation model by annotating CholecT50 and Cholec80 with their corresponding tool masks in order to build the training dataset for our ControlNet, discarding frames that do not contain tools (yielding 84,857 samples), and to create conditioning videos for ControlVideo.

\subsection{Training and Inference Details}
The models are trained on three NVIDIA A100 GPUs with a per-GPU batch size of 128 for StableDiffusion finetuning and a per-GPU batch size of 256 for ControlNet training. We use \texttt{pytorch} along with the Hugginface libraries \texttt{diffusers}, which provides implementations for diffusion models, and \texttt{accelerate}, which enables distributed training.
In both training stages 1 and 2, we use a constant learning rate of 1e-5.
For StableDiffusion finetuning (Stage 1), we have two slightly different training setups: While one setup utilizes a frozen CLIP text encoder, the alternative setup trains both CLIP text encoder and UNet. The latter allows our StabelDiffusion model to adapt to the laparoscopic domain with particular text prompts. The training leads to sufficient image quality after 10 epochs, which takes up to one hour on our training setup.
ControlNet (Stage 2) is trained for 100 epochs taking around 2.5 hours. ControlVideo Inference (Stage 3) for a 10-second video at 25fps takes roughly 1.5 minutes. All images and videos are generated with 100 inference steps and a guidance scale of 3.

\begin{table}[t]
\centering
\begin{tabular}{l|c||c|c|c|c||c}
\textbf{Model / Setup} & \textbf{Epochs} & \textbf{CMMD$\downarrow$} & \textbf{CLIP-FID$\downarrow$} & \textbf{FID$\downarrow$} & \textbf{KID$\downarrow$} & \textbf{action mAP$\uparrow$} \\
\hline
\multirow{3}{*}{\shortstack{CholecT50\\(UNet)}} & 1 & 0.206 & 4.991 & 50.637 & 0.04359 & 0.0421\\
& 5 & 0.103 & 3.42 & 36.962 & 0.02660 & 0.0971\\
& 10 & 0.113 & 3.522 & 38.548 & 0.02953 & 0.1325\\
\hline
\multirow{3}{*}{\shortstack{CholecT50+80\\(UNet)}} & 1 & 0.174 & 4.32 & 49.428 & 0.04311 & 0.0526\\
& 5 & 0.113 & 3.531 & 40.011 & 0.03024 & 0.0939\\
& 10 & \textbf{0.102} & \textbf{3.298} & \textbf{33.426} & \textbf{0.02203} & 0.1457\\
\hline
\multirow{3}{*}{\shortstack{CholecT50+80\\(UNet+CLIP)}} & 1 & 0.14 & 3.804 & 44.045 & 0.03616 & 0.0503\\
& 5 & 0.103 & 3.337 & 38.331 & 0.02858 & 0.1035\\
& 10 & 0.119 & 3.37 & 34.197 & 0.02457 & \textbf{0.2055} \\
\hline
\hline
ControlNet & 100 & 0.124 & 2.474 & 38.097 & 0.02924 & 0.2516\\
\textit{CholecT50 (RDV)} & - & - & - & - & - & 0.2990
\end{tabular}
\caption{Quantitative assessment of fidelity for the two different dataset variants across epochs. The third setup (UNet+CLIP) finetunes both UNet and the CLIP text encoder, while the other two keep CLIP frozen. We use the text-to-image model with the best surgical action mAP for the ControlNet training. The ControlNet further improves factual correctness while preserving good fidelity metrics. For comparison, we also report the action recognition performance on actual data from the CholecT50 (\textit{CholecT50 (RDV)}).}
\label{tab:fid}
\vspace{-10px}
\end{table}

\subsection{Evaluation}
For \textbf{StableDiffusion finetuning (Stage 1)}, we evaluate our model on the fidelity and factual correctness of the resulting generated images. For this purpose, we generate 10,000 images with random text prompts for each dataset variant and setup. These generated images are assessed with the same number of real images with equal text prompts from the CholecT50 dataset exemplary displayed in \Cref{fig:examples}.
As all of the common fidelity metrics build upon feature extractor models that were trained on non-medical images, we use four different metrics to better account for variances with regard to the specific domain of laparoscopic surgery.
The Fréchet Inception Distance (\textbf{FID})~\cite{FID} and the Kernel Inception Distance (\textbf{KID})~\cite{KID} use an InceptionV3 model trained on ImageNet, while the variant \textbf{CLIP-FID}~\cite{cleanfid} and the recently proposed \textbf{CMMD}~\cite{cmmd} use the image encoder from the pre-trained vision-transformer model CLIP-ViT-B-32 \cite{clip}. Although the absolute scores of the fidelity metrics vary, the relative tendency of the best configuration is similar, proving the robustness of our result (\Cref{tab:fid}). Factual correctness is evaluated using Rendezvouz~\cite{rendezvous} (RDV), a surgical action recognition model that was introduced in combination with the CholecT50 dataset. We report the average mAP of all possible surgical action triplets $<$\textit{tool, verb, target}$>$. 

\begin{figure}[t]
    \centering Prompt: \textbf{"grasper grasp gallbladder in carlot-triangle dissection"}\\
    \vspace{2px}
    \centering\includegraphics[scale=.35]{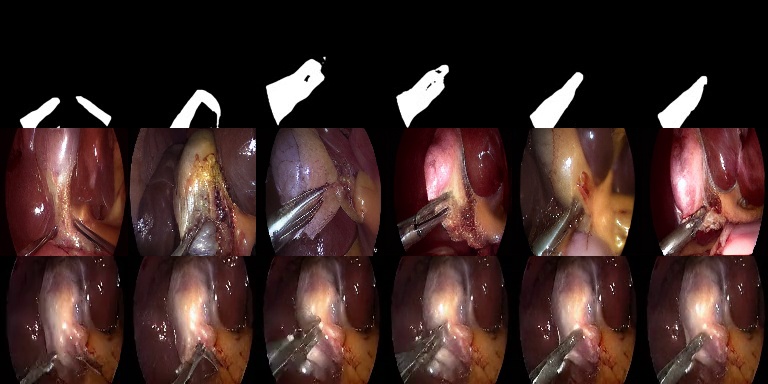}\vspace{10pt}\\
    \centering Prompt: \textbf{"grasper retract liver and irrigator aspirate fluid in cleaning-and-coagulation"}\\
    \vspace{2px}
    \centering\includegraphics[scale=.35]{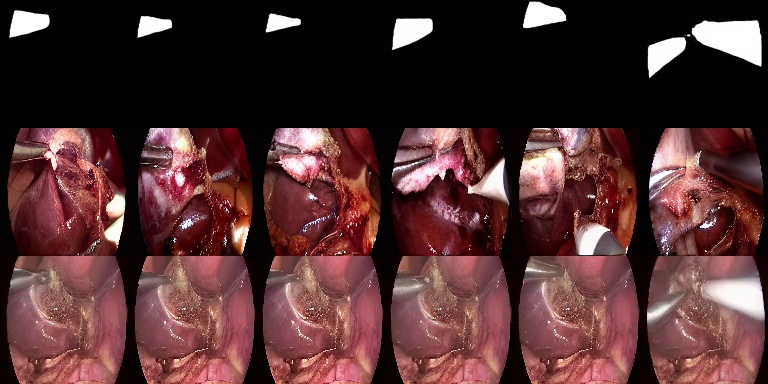}
    \caption{Generated images/video-frames for the given text prompts and tool masks. Displayed w.r.t the temporal dimension from left to right. \textbf{First row:} tool conditioning (input), \textbf{second row:} ControlNet output, \textbf{third row:} ControlVideo output}
    \label{fig:controlvideo}
    \vspace{-10px}
\end{figure}

Our findings from \Cref{tab:fid} indicate that using the datasets CholecT50 and Cholec80, as well as finetuning the CLIP text encoder (10 epochs), achieves the best result with good fidelity and the best overall factual correctness scores (action mAP). This shows that finetuning the text encoder helps StableDiffusion to adapt to the laparoscopic domain. This resulting model is used to train ControlNet in Stage 2.

The evaluation of the \textbf{ControlNet training (Stage 2)} assesses the ability of the model to generate tools with spatial precision. Using our finetuned YOLOv8 model, we predict tool segmentation masks of 4000 images, which are generated from masks unseen during training. The masks are evaluated by calculating the pixel-wise F1 score between input conditioning and prediction. Our model achieves an F1 score of 0.71 and  improves factual correctness of the generated images, while preserving good fidelity metrics. 

As \textbf{ControlVideo Inference (Stage 3)} is a zero-shot method directly utilizing our StableDiffusion and ControlNet models with no additional training, we just provide a qualitative analysis of the generated videos.

\section{Discussion and Conclusion}
In this paper, we approach the generation of laparoscopic simulations leveraging the open-source models StableDiffusion for text-to-image generation, ControlNet for adding tool position conditioning, and ControlVideo to interactively produce videos from text prompts and tool movements.
Our results demonstrate state-of-the-art fidelity with an FID score of $33.43$ as well as a pixel-wise F1 score of $0.71$ for the control of tool positions. We show coherent and realistic-looking videos with a successful generation of surgical actions. These advancements mark a significant step forward in the realm of surgical simulations, offering a promising avenue for enhanced medical education and training.

However, despite the notable progress, our evaluation reveals certain limitations. While our approach already demonstrates some preservation of factual correctness, there remains a gap to actual surgical videos that needs to be bridged. Although our ControlNet effectively generates tools on the image with reasonable spatial precision, but there is room for improvement of the spatial conditioning, as indicated by the F1-score. Further iterations of our approach could benefit from more detailed tool positions in 3D (e.g., in a virtual reality setting), such as depth maps, to enhance the tool movements even more. Additionally, extended conditioning considering camera movements could add to realism and overcome the currently quite static backgrounds. We also recognize the potential for training-based video generation methods to capture the semantics of the surgical actions more effectively in subsequent frames.

In conclusion, our work represents a notable effort in the domain of controllable surgical video generation. By addressing the identified limitations and incorporating feedback from domain experts, we anticipate continued progress toward more immersive and effective surgical training platforms, ultimately benefiting medical professionals and patients alike. A promising field of research lies ahead.
%
%
%
\bibliographystyle{splncs04}
\bibliography{references}

\end{document}